\documentclass[conference]{IEEEtran} 
\IEEEoverridecommandlockouts
\usepackage{setspace}
\usepackage{subfigure}
\usepackage{graphicx}
\usepackage{epstopdf}
\usepackage{float}
\usepackage{amsmath}
\usepackage{amsmath,amssymb,amsfonts}
\usepackage[ruled]{algorithm2e}
\usepackage{algpseudocode}
\usepackage{array}
\usepackage{amsthm}
\usepackage{amsmath}
\usepackage{amssymb}
\usepackage{mdwmath}
\usepackage{mdwtab}
\usepackage{eqparbox}
\usepackage{stfloats}
\usepackage{fixltx2e}
\usepackage{cases} 
\usepackage{caption}
\usepackage{graphicx}
\usepackage{float} 

\usepackage{flushend}

\usepackage{xcolor}
\usepackage{makecell}

\usepackage{url}
\usepackage{textcomp}
\usepackage{cite}
\def\BibTeX{{\rm B\kern-.05em{\sc i\kern-.025em b}\kern-.08em
    T\kern-.1667em\lower.7ex\hbox{E}\kern-.125emX}}
\allowdisplaybreaks[4]

\usepackage{wrapfig}

\makeatletter
\renewcommand{\maketag@@@}[1]{\hbox{\m@th\normalsize\normalfont#1}}%
\makeatother
\UseRawInputEncoding 

\makeatletter
\newcommand{\linebreakand}{%
  \end{@IEEEauthorhalign}
  \hfill\mbox{}\par
  \mbox{}\hfill\begin{@IEEEauthorhalign}
}
\makeatother

\begin{document}

\title
{Near-Field Integrated Sensing and Communications:  Unlocking Potentials and Shaping the Future}

\author{\IEEEauthorblockN{Kaiqian Qu, Shuaishuai Guo}
\IEEEauthorblockA{\textit{Control Science and Engineering,}\\\textit{Shandong University} \\
Email: qukaiqian@mail.sdu.edu.cn\\ shuaishuai\_guo@sdu.edu.cn}
\and

\IEEEauthorblockN{ Jia Ye}
\IEEEauthorblockA{\textit{School of Electrical Engineering,}\\\textit{Chongqing University} \\
Email: yejiaft@163.com}

\and

\IEEEauthorblockN{ Nasir Saeed}
\IEEEauthorblockA{\textit{Department of Electrical and}\\ \textit{ Communication Engineering,}\\\textit{ United Arab Emirates University}\\
Email: mr.nasir.saeed@ieee.org
}
}

\maketitle


\begin{abstract}
The sixth generation (6G) communication networks are featured by integrated sensing and communications (ISAC), revolutionizing base stations (BSs) and terminals. Additionally, in the unfolding 6G landscape, a pivotal physical layer technology, the Extremely Large-Scale Antenna Array (ELAA), assumes center stage. With its expansive coverage of the near-field region, ELAA's electromagnetic (EM) waves manifest captivating spherical wave properties. Embracing these distinctive features, communication and sensing capabilities scale unprecedented heights.  Therefore, we systematically explore the prodigious potential of near-field ISAC technology. In particular, the fundamental principles of near-field are presented to unearth its benefits in both communication and sensing. Then, we delve into the technologies underpinning near-field communication and sensing, unraveling possibilities discussed in recent works. We then investigated the advantages of near-field ISAC through rigorous case simulations, showcasing the benefits of near-field ISAC and reinforcing its stature as a transformative paradigm.
As we conclude, we confront the open frontiers and chart the future directions for near-field ISAC.




\end{abstract}


\section{Introduction} 
The sixth generation (6G) networks are set to enable cutting-edge applications and the Internet of Everything (IoE). With extremely low latency ($0.1$ ms) and high transmission rates ($1$ Tbps), 6G opens up potential scenarios like digital twins, smart cities, and smart homes. To achieve these goals, research focuses on technologies like Integrated Sensing and Communications (ISAC), Extremely Large-Scale Array (ELAA), and Reconfigurable Intelligent Surface (RIS) \cite{Jia2020}.Among them, ISAC represents a transformative field at the intersection of sensing technology and wireless communications.
The convergence arises from the striking similarity in hardware structures, with communication and radar systems employing radio frequency (RF) chains and transceivers.
 On the other hand, the resurgence of ISAC technology is prompted by the scarcity of spectrum resources due to the increasing demand for wireless communication services. ISAC can lead to the overlapping of communication frequency bands with radar systems, creating opportunities for higher spectrum efficiency \cite{10437489,10623740}.

Advancements in antenna structures like ELAA and RIS are crucial for 6G, offering increased beamforming gain facilitated by a higher number of antennas or reflecting elements. These technologies change the system's electromagnetic (EM) properties, with the Rayleigh distance, dependent on array aperture and frequency, marking the boundary between near-field and far-field regions. Traditional millimeter-wave MIMO systems, with fewer antennas, assume far-field conditions where EM waves are plane waves. In contrast, ELAA and RIS can extend the Rayleigh distance to hundreds or thousands of meters, signifying a substantial transition to the near-field regime. In the near field, EM waves exhibit distinctive spherical wave characteristics with complex and spatially varying propagation patterns. The spatial dispersion presents both challenges and opportunities for designing and optimizing 6G systems.  
Understanding and effectively managing these near-field effects are critical in fully harnessing the potential of ELAA and RIS technologies to maximize their benefits in the next generation of communication systems.

Spherical wavefronts can be effectively utilized to generate highly focused beams within specific spatial regions, a phenomenon known as beam focusing. By exploiting spherical wavefronts, it becomes possible to focus the energy on desired areas, enabling precise localization and enhanced performance in targeted regions. Unlike traditional far-field beam steering, where signals can only be directed toward a specific direction, this approach provides unprecedented control and achieves superior beamfocusing capabilities. 
The existing studies indicate that near-field beamfocusing (or spherical wave) can help decorrelate muti-user channel \cite{sperial}, increase spatial degrees of freedom (DoF) \cite{lu,10262267}, and provide new multi-access methods\cite{ldma}. In addition, angle and distance information is carried in the spherical wave, providing additional insights for target localization\cite{9508850}.
 As a result, the distinctive properties of the near field hold the promise of enhancing communication and sensing performance, opening up new and exciting opportunities for ISAC \cite{Wang2023NearFieldIS}. However, despite its potential, near-field ISAC remains relatively unexplored, prompting us to embark on a systematic exploration of its capabilities.

This article systematically introduces near-field ISAC technology. Key features include:
\begin{itemize}
    \item \textbf{Advancing Near-Field Understanding:} We explore the unique characteristics of near-field spherical waves and their benefits for communication and sensing, laying the groundwork for near-field ISAC systems.
    \item \textbf{Empowering Near-Field ISAC Applications:} We provide an overview of near-field communication and sensing, analyzing practical applications and demonstrating the transformative impact of near-field techniques.
    \item \textbf{Unlocking Potentials and Shaping Future:} We identify critical research challenges, explore the compatibility of near-field EM models with existing frameworks, and propose improvements to redefine near-field distances for more accurate ISAC systems. These insights aim to inspire future research in this dynamic field.
\end{itemize}

\section{Near-field Communications and Sensing}
In this section, we present the foundations of near-field, the benefits of near-field for communication and sensing, and a discussion on technologies to be revisited in near-field ISAC.
\subsection{What is Near-field?}
\begin{table*}[htbp]
       \centering
       \includegraphics[width=0.9\linewidth]{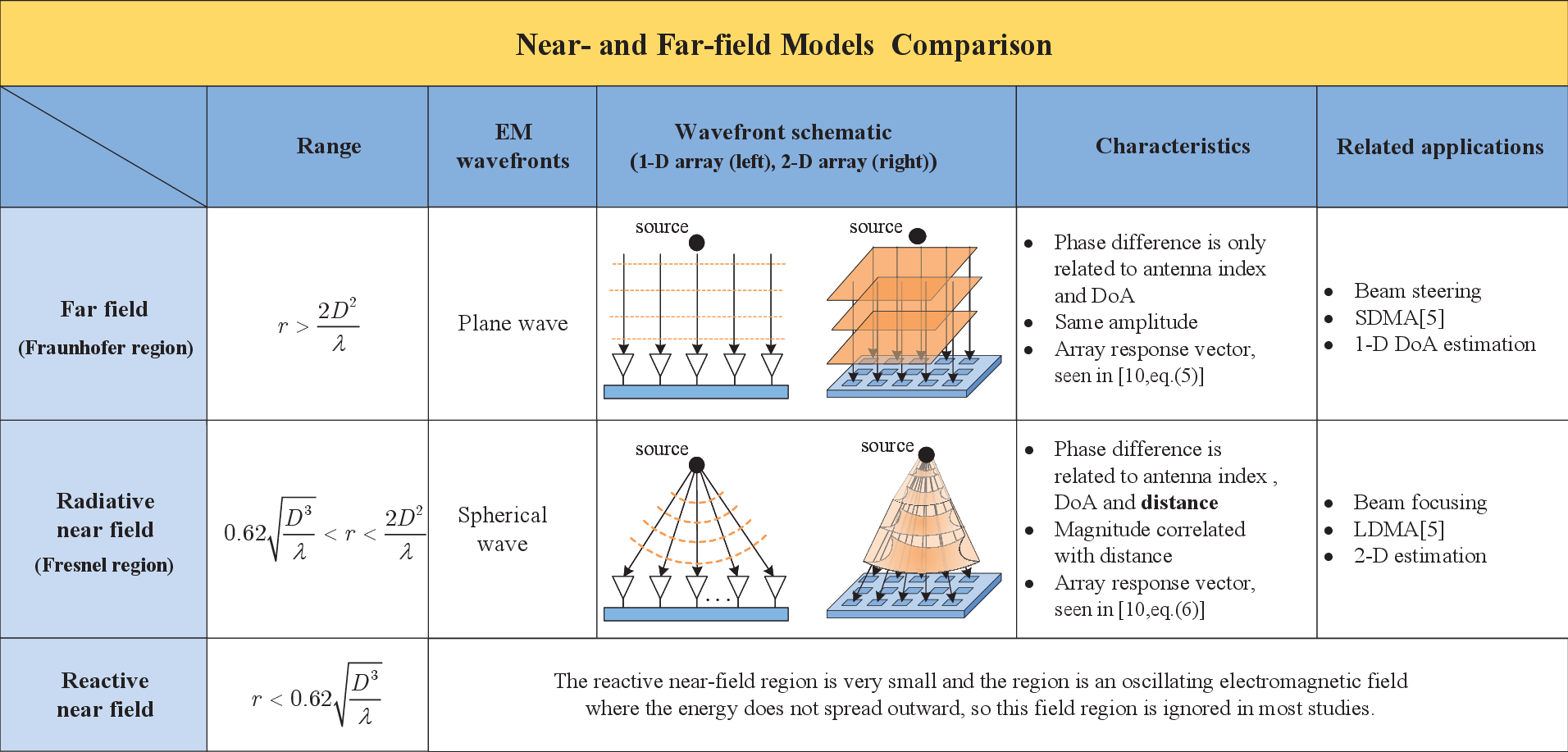}
       \caption{Near- and Far-field Models Comparison. The following abbreviations are used for brevity: direction of arrival (DoA),  spatial division multiple access (SDMA), location division multiple access (LDMA).}
       \label{Model}
\end{table*}
In EMs and acoustics, near-field and far-field are two regions that describe the distance relationship to a radiation source. 
When signals propagate in free space, they spread out in a spherical pattern. In the far-field region known as the radiation zone or Fraunhofer region, at the distance the distance from the source of the wavefront is much greater than its size, the curvature of the spherical wave is small and can be approximated as a plane wave. 
However, in the near-field region known as the reactive or Fresnel region, the distance from the source of the wavefront is comparable to or smaller than its size, the wavefront remains distinctly spherical, as illustrated in Table \ref{Model}.
In reality, there is no strict distance threshold between the near and far fields; it must be determined based on specific applications and scenarios. 
The most commonly used criterion to distinguish between the far-field and near-field regions is the Rayleigh distance, which accounts for the phase difference caused by the curvature of the EM wave between the center and boundary of the receiving array.
 When the phase difference is less than 22.5 degrees, it is considered a small curvature, and the wavefront can be approximated as a plane wave; 
Otherwise, it keeps the spherical wave. Mathematically, as defined in Table \ref{Model}, the Rayleigh distance is proportional to the product of the carrier frequency $\frac{1}{\lambda}$ and the square of the array aperture size, $D^2$.

Unlike the far-field plane wave, where the wavefront's behavior is primarily determined by the angle of propagation, near-field introduces an additional distance dimension, playing a significant role in determining the wavefront's shape and behavior. 

\subsection{Benefits of Near-field for Communication and Sensing}
\subsubsection{\textbf{For Communication}}
On one hand, the characteristics of near-field spherical waves can increase the DoFs of the channel and improve the system capacity \cite{10129110}. In a typical far-field millimeter-wave point-to-point communication system with $N_t$ transmitting antennas and $N_r$ receiving antennas, the phase difference between the transceiver antenna is linear. Therefore, the line of sight (LoS) channel is simplified as the product of the steering vector with only one DoF. In the near field, a big difference is that the phase between the transceiver antennas is nonlinear, which is related to the distance of each antenna. In other words, in the far field, all antennas at the receiving or transmitting end are assumed to share a single distance value.  
However, in the near field, each antenna has its own distance information\footnote{This is related to the center distance between the transceiver. When the distance is small enough, the distance between each antenna varies significantly, providing the highest DoFs. As the distance increases slightly, several adjacent antennas share distance information and the DoF gradually decreases until it drops to 1 in the far field.}, and the DoF is equal to $\min\{N_t,N_r\}$, as shown in Fig. 1. Similarly, the above analysis still holds in multiuser MIMO. Moreover, in the far field, users located at the same or similar angles cannot be distinguished, leading to a decrease in the DoFs. Whereas in the near field, these users can be distinguished by different distances, effectively improving the DoFs.

On the other hand, the additional distance dimension in the near field introduces a new dimension to mitigate multi-user interference. In the traditional far-field communication model, it is not possible to differentiate between users with the same angle, leading to significant inter-user interference. In contrast, the near field can distinguish users based on different distances and effectively mitigate inter-user interference. 
In addition, a similar benefit is that the spherical wave helps to decorrelate the multi-user channel to make it close to the optimal propagation condition, as shown in Fig. 2.

 
\subsubsection{\textbf{For Sensing}}
The process of sensing can be divided into signal transmission and reception. 
 In near-field signal transmission, the use of focused and tightly confined beams is possible. This allows for focusing the signal energy on the target point with reduced signal spreading, leading to improved signal strength at the target.

In the stage of signal reception processing, near-field spherical waves carry both distance and angle information. 
In addition, near-field reception offers the advantage of selective signal capture, enabling targeted reception from specific transmitters and minimizing the impact of unwanted signals or interference. To derive precise position information of the target, near-field reception can be combined with advanced estimation techniques and spectral search algorithms, resulting in significantly higher accuracy compared to the far-field scenario.
\begin{figure}[t]
       \centering
       \includegraphics[width=0.9\linewidth]{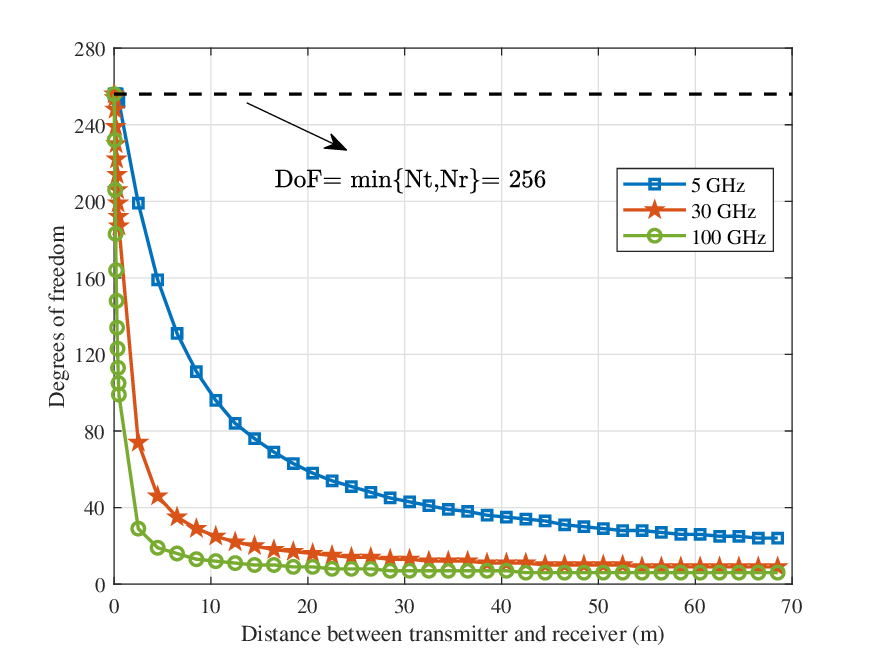}
       \caption{The spatial degrees of freedom of point-to-point MIMO system varies with the distance between transmitter and receiver, $N_t=N_r=256$.}
       \label{fig1}
\end{figure}

\begin{figure}[t]
       \centering
       \includegraphics[width=0.9\linewidth]{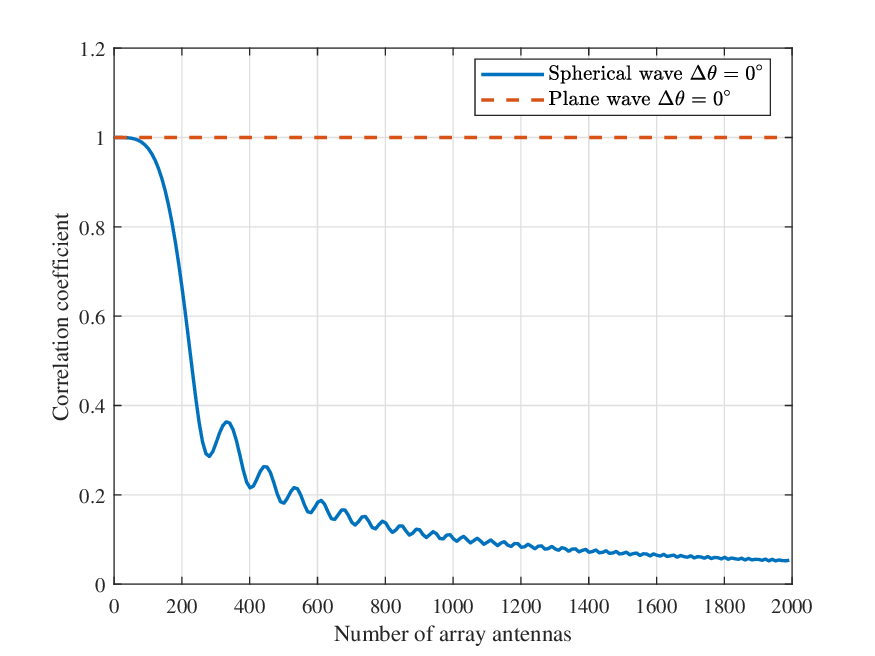}
       \caption{Squared-correlation coefficient versus antenna number for the Plane and Spherical wave models. The two users are located in the same direction.}
       \label{fig2}
\end{figure}

\subsection{Technologies to be Revisited in Near-Field ISAC}
The emergence of the near field has attracted the attention of researchers in widespread fields. In this subsection, we briefly introduce the communication and sensing technologies to be revisited in the near field.

\subsubsection{\textbf{Near-field Channel Estimation}}
 Channel estimation plays a pivotal role in enhancing the performance of ISAC systems. In fact, the target sensing process in ISAC networks can be likened to channel estimation in communication aspects, with the distinction that sensing operates in a backscatter channel. Accurate channel estimation in ISAC systems, therefore, yields advantages for both communication and sensing domains. It enables the deployment of advanced signal processing techniques, enhances spectral efficiency, improves localization accuracy, and fosters more reliable and efficient communication and sensing operations.

The structural changes in near-field EM waves caused by large-scale arrays render conventional channel estimation methods ineffective for near-field channel estimation. 
In \cite{8949454}, the authors uniformly divided the two-dimensional physical space into multiple grids, each corresponding to a near-field array response vector. These near-field response vectors form a codebook for compressive sensing-based recovery of channel information. However, in recent studies, the orthogonality of near-field polar regions has been demonstrated \cite{9693928}. Since the correlation of the near-field beam varies non-uniformly along the distance dimension, a polar domain non-uniform sampling codebook is designed in \cite{9693928}, which can match the near-field channel well.

\subsubsection{\textbf{Near-Field Multiple Access}}
Spatial division multiple access (SDMA) is a key technology used to enhance MIMO communication's spectrum efficiency. 
In ISAC networks, SDMA can be particularly beneficial in scenarios where communication and sensing tasks need to be performed simultaneously or in close temporal proximity. SDMA's capability to create spatially distinct channels allows for simultaneous and independent signals transmission to multiple users and targets, providing a valuable solution for effectively managing communication and sensing tasks in close temporal proximity. 
ELAA introduces an additional resolution in the near-field domain, allowing spatial resources to be further divided into different grids based on distance and angle. 
According to the characteristics of the near field, location division multiple access (LDMA)  \cite{ldma} is a promising near-field multiple access technology. The core idea is to utilize additional spatial resources in the distance domain to serve different users in different positions in the near field (determined by angle and distance).
\begin{figure*}[t]
       \centering
       \includegraphics[width=0.9\linewidth]{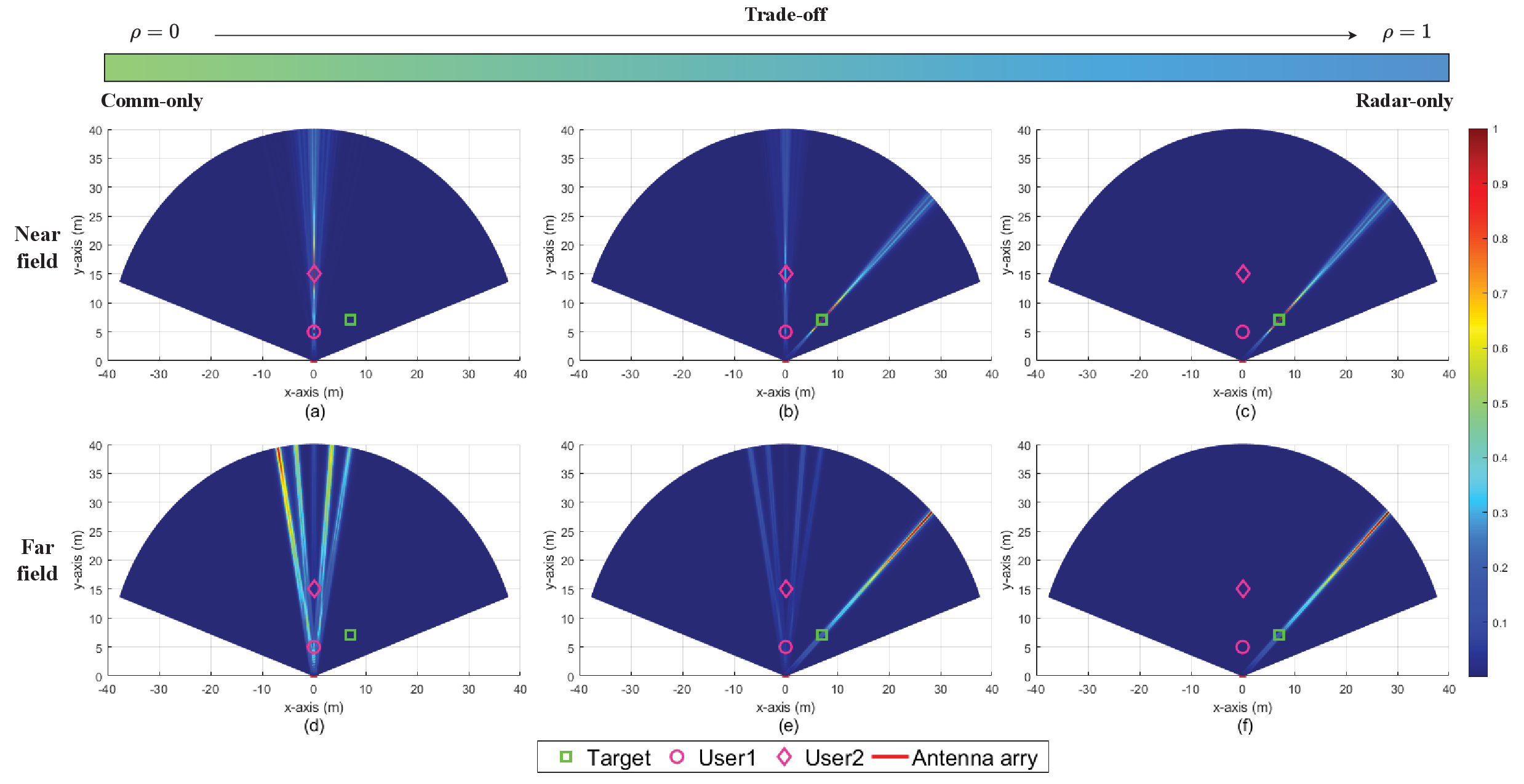}
       \caption{The normalized signal power measurement of beams, (a) Comm-only NFBF; (b) Trade-off $\rho=0.5$ NFBF; (c) Radar-only NFBF; (d) Comm-only FFBF; (e) Trade-off $\rho=0.5$ FFBF; (f) Radar-only FFBF; }
       \label{fig5}
\end{figure*}
\subsubsection{\textbf{Near-field Positioning Technology}}
Source localization and tracking in ISAC networks rely on the joint estimation of the angle of arrival (AoA) and the time of arrival (ToA). These functionalities are crucial for applications like indoor localization, smart environments, wireless sensor networks, and surveillance systems. However, they require good synchronization between transmitters and receivers or the involvement of multiple nodes, which can be limiting.
Fortunately, near-field propagation offers advantages that enhance the performance of source localization and tracking in ISAC networks. When the antenna array is large enough, electromagnetic waves exhibit a spherical wave shape in the near field of an extremely large antenna array (ELAA). A promising approach is to directly extract the source position from the spherical wavefront impinging on a single large array, eliminating the need for synchronization \cite{9508850}. This insight into positioning technology from near-field spherical waves may also lead to holographic positioning.

\subsubsection{\textbf{Near-field Signal Processing} }
Signal processing techniques in ISAC networks are essential for efficient communication, accurate source localization, and robust interference mitigation. Two fundamental tasks are DoA estimation and beamforming. DoA estimation determines the angles at which signals arrive at the antenna array, while beamforming directs transmitted or received signals towards specific positions and suppresses interference from other directions. These tasks are interrelated, as accurate beamforming relies on precise DoA acquisition and vice versa.
However, conventional beamforming used in far-field scenarios cannot be directly applied to near-field environments due to complex spatial variations like varying signal strengths and phase shifts across the antenna array. This mismatch results in significant performance loss and prevents effective DoA estimation. In the near-field, far-field beams become divergent and wider, leading to increased user interference and angle estimation errors, diminishing communication rates and sensing accuracy. To fully leverage near-field benefits in ISAC networks, specialized beamforming techniques and advanced DoA estimation algorithms are necessary \cite{9723331,8359308}. These approaches aim to address the challenges posed by near-field propagation characteristics and the trade-offs between communication performance and sensing accuracy.

\section{Experiment and Discussion}
To demonstrate the performance and potential of near-field ISAC, we conducted several numerical experiments. In our scenario, a base station (BS) is equipped with uniform linear arrays (ULA) with 256 transmitting and 256 receiving antennas, serving two single-antenna users and performing a single-target sensing task, all in the near field. The users are located at the same angle with coordinates $(0^\circ, 5~m)$ and $(0^\circ, 15~m)$, and the target is at $(45^\circ, 5~m)$. The system operates at $30$ GHz, with each user channel containing two scattering paths and one LoS path.

We evaluated both near-field beamforming (NFBF) and far-field beamforming (FFBF) designs\footnote{Note that zero-forcing (ZF) precoding is adopted for the communication part of ISAC beamforming.}.
Fig. 3 shows the effect of integrated beam design with different weighting factors. Fig. 3(a), (b), and (c) demonstrate excellent beam focusing on both users and the target, effectively managing inter-user interference. Figures 3(d), (e), and (f) show that traditional far-field beamforming suffers from significant performance loss due to mismatch, failing to distinguish between users and experiencing severe gain loss.
Notably, FFBF directs beam energy more toward the scattering path rather than the LoS path due to ZF precoding, which relies on the scattering path to distinguish users and reduce interference. This means FFBF's communication performance is limited by the number of scattering paths, impacting efficiency in environments with few or poorly distributed scattering paths.

Next, we analyze the trade-off performance between communication and sensing under the NFBF and FFBF designs in Fig. 4, where the sensing performance is evaluated via the root Cramer-Rao bound (RCRB). The curve clearly shows the existence of a trade-off, regardless of whether NFBF or FFBF is used.
The communication rate of FFBF quickly saturates around 5 bit/Hz/s, while NFBF reaches approximately 24 bit/Hz/s. This is because FFBF can only use the scattering path to distinguish between users, and suffers from beamforming gain loss due to the inaccurate plane wave assumption.
Furthermore, the estimation performance of FFBF deteriorates with decreasing target distance, unlike NFBF. The degradation is due to the greater gain loss experienced by FFBF in the near field, resulting in lower power at the target and poorer estimation performance. Overall, the analysis indicates that NFBF exhibits a better trade-off performance.

\begin{figure}[t]
       \centering
       \includegraphics[width=0.9\linewidth]{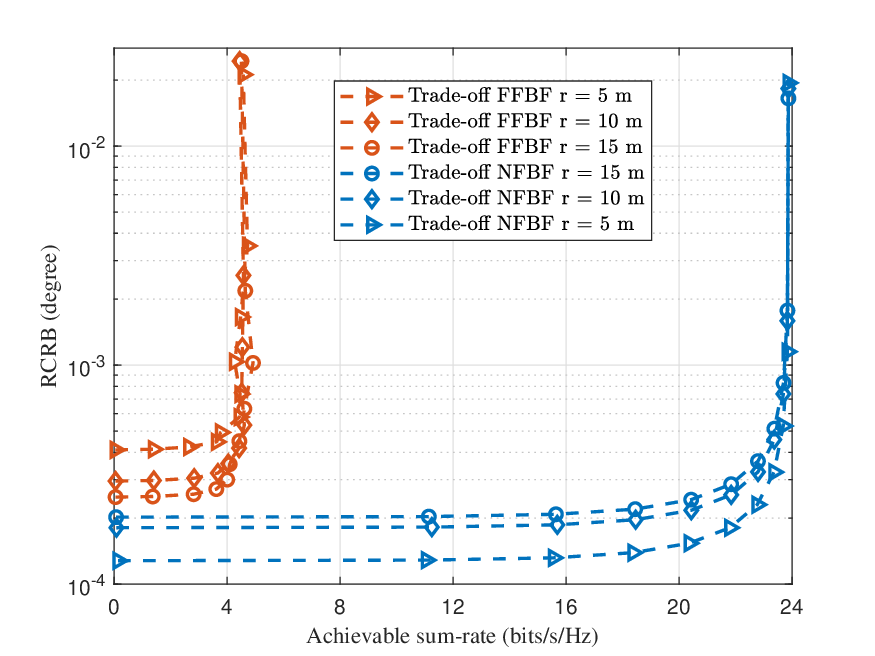}
       \caption{RCRB versus communication rate, a single target is located at different distance $r$.}
       \label{fig3}
\end{figure}
\begin{figure}[t]
       \centering
       \includegraphics[width=0.9\linewidth]{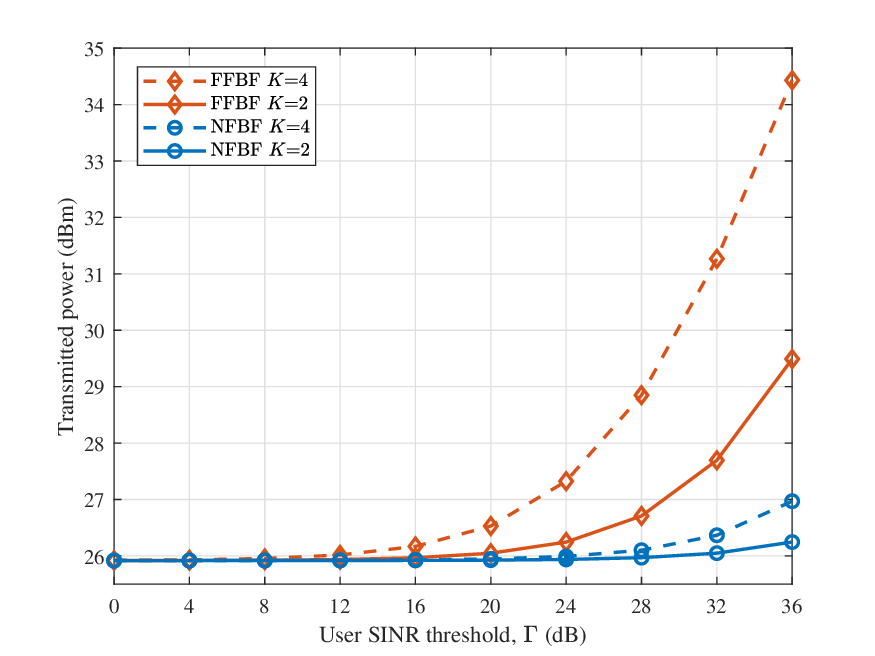}
       \caption{Transmitted power varies with the SINR threshold of the user.}
       \label{fig4}
\end{figure}

In Fig. 5, we aim to reveal the power-saving potential of near-field ISAC. 
Since the estimation performance of the target is positively correlated with the signal power at the target location, we compare the required minimum transmit power under the constraint of satisfying the communication SINR threshold and maintaining a fixed power at the target. 
 Clearly, as the SINR threshold increases, FFBF requires more power than NFBF to satisfy the SINR. This observation aligns with the previous analysis, where FFBF relies on scattering paths with higher pathloss, necessitating higher power to meet the SINR requirements.
 Furthermore, a noteworthy observation arises in the context of near-field beam focusing: the closer the distance between BS and users, the more pronounced the focusing effect becomes. This compelling finding suggests a hypothesis that the communication rate in near-field ISAC will exhibit an upward trend as the distance decreases. The increase in communication rate is owing to the enhanced beam focusing effect, which allows for more precise differentiation of nearby users and consequently leads to improved interference reduction.

 In summary, the comprehensive analysis and results presented highlight the significant advantages of implementing near-field ISAC in future 6G large-scale antenna systems.

\section{Open Challenges and Future Directions} 
In summary, beam focusing plays a crucial role in enhancing near-field ISAC performance and holds promising potential for energy conservation. Nevertheless, despite its benefits, there remain several potential challenges and avenues for further research in the field of near-field ISAC, which are briefly discussed in this section.
\subsection{Accurate Near-field Model} 
The significance of near-field models in contrast to far-field models has been demonstrated, highlighting their critical role in minimizing gain and accuracy losses in near-field conditions. However, it is essential to determine if existing mathematical models truly capture authentic near-field characteristics. Recent research \cite{8736783} suggests that current near-field models may not accurately represent actual near-field EM behaviors. These models often assume a near-field identical to the far-field, merely correcting for spherical propagation curvature, and overlook range-dependent amplitude variations.
Given these findings, researchers must acknowledge the limitations of current near-field models and explore new approaches that better capture the complexities of real-world near-field EM behaviors. This could lead to more accurate and reliable results in practical applications.

\subsection{Near-field Distance Improvement}
The distinction between the near and far fields is not strictly confined, and the Rayleigh distance, based solely on phase difference, fails to encompass all performance metrics comprehensively \cite{9508850}. Take, for instance, the 256-256 parallel-placed ULA system illustrated in Fig. 1, where the spatial DoF experiences gradual growth only within distances less than 70 meters, deviating significantly from the Rayleigh distance. Conversely, the near-field distance serves as a valuable reference and can serve as a decisive factor for employing near-field intelligent spectrum access and control techniques when the distance is below this threshold. As such, it becomes essential to devise near-field distances that capture various performance aspects, including the overall effectiveness of ISAC.


\subsection{Near-field Wideband ISAC}. 
Wideband ISAC holds great promise, especially in the Terahertz (THz) band. However, it faces challenges due to beam splitting in wideband scenarios. In near-field broadband systems, conventional frequency-independent phase shifter beamforming results in distinct focusing points at different frequencies, and this effect extends to different directions in the far field as well \cite{cui2021nearfield}. Such beam splitting hinders the generation of accurate broadband beams, thereby failing to fully exploit the gain potential offered by high bandwidth. To address this issue, a phase-delay focusing (PDF) method has been proposed in \cite{cui2021nearfield}, leveraging true time delay (TTD)-based beamforming techniques. Furthermore, it's important to note that the wideband DoA estimation methods used in the far field are not directly applicable to near-field wideband ISAC scenarios, necessitating the design of specialized algorithms for this purpose. Consequently, near-field ISAC for high-frequency wideband operations represents a worthwhile and significant research problem that merits attention and exploration.

\subsection{Distributed XL-MIMO Near Field}
Distributed MIMO can achieve a distributed large aperture through different nodes. Compared to centralized MIMO, its virtual aperture is larger and more uniform. This results in more stable near-field effects, such as more precise beam focusing. Additionally, existing cooperative multi-point transmission and distributed MIMO radar technologies can already form a preliminary distributed MIMO ISAC. However, to fully leverage the advantages of distributed apertures, synchronization between different nodes is a practical challenge.

\section{Conclusions}The emergence of ELAA as a prevailing trend in 6G base stations has elevated the significance of the near-field region, where EM waves exhibit spherical wavefronts, demanding utmost attention and consideration. In light of the 6G vision, ISAC must also account for the near-field impact. This paper has explored the potential of near-field ISAC techniques. Initially, we elucidated the fundamental aspects of the near field and compared its characteristics with those of the far field. Subsequently, we highlighted the numerous advantages of the near field for communication and sensing, accompanied by an in-depth examination of relevant technological studies.
Moreover, our numerical results showcased the superior trade-off performance achieved through near-field ISAC designs, emphasizing the potential for power-saving advantages. As we conclude, we underscored some challenges that need to be addressed and identified promising avenues for future research in this domain.
\bibliographystyle{IEEEtran} 
\bibliography{IEEEabrv,bib}
\end{document}